\documentclass[times,authoryear,3p]{elsarticle}
\usepackage{graphicx,amssymb,color}
\def\arcsec{\hbox{$^{\prime\prime}$}}
\journal{Icarus}
\begin{document}
\begin{frontmatter}
\title{Stability analysis of the Martian obliquity during the Noachian era}
\author{Ramon Brasser\corref{ca}}
\cortext[ca]{Corresponding author}
\ead{brasser\_astro@yahoo.com}
\address{Dep. Cassiop\'{e}e, University of Nice - Sophia Antipolis, CNRS, Obs. de la C\^{o}te d'Azur; Nice, France}
\author{Kevin J. Walsh}
\ead{kwalsh@boulder.swri.edu}
\address{Dep. Cassiop\'{e}e, University of Nice - Sophia Antipolis, CNRS, Obs. de la C\^{o}te d'Azur; Nice, France}
\address{Southwest Research Institute, 1050 Walnut Street, Suite 300; Boulder, CO 80302, USA}
\begin{abstract}
We performed numerical simulations of the obliquity evolution of Mars during the Noachian era, at which time the giant planets were on
drastically different orbits than today. For the preferred primordial configuration of the planets we find that there are two large
zones where the Martian obliquity is stable and oscillates with an amplitude lower than 20$^\circ$. These zones occur at
obliquities below 30$^\circ$ and above 60$^\circ$; intermediate values show either resonant or chaotic behaviour depending on the
primordial orbits of the terrestrial planets.
\end{abstract}
\end{frontmatter}

\section{Introduction}
The current obliquity of Mars is chaotic with a Lyapunov time scale of 5~Myr (Laskar \& Robutel, 1993; Touma \& Wisdom, 1993). This
has drastic consequences for the Martian climate. At present, the obliquity is a modest 25$^\circ$, the average polar insolation is
lower than the equatorial one (Ward, 1974) and ice caps are observed at the poles of the planet. However, during phases where
the obliquity is high, $\varepsilon \gtrsim 40^\circ$, large quantities of polar ice sublimate and could be transported to the tropical
regions (Jakosky \& Carr, 1985; Jakosky et al., 1995). On shorter time scales the Martian obliquity suffers large-amplitude
oscillations of approximately 22$^\circ$ around a mean value of 25$^\circ$ with a period of 125~kyr (Ward, 1974), caused by the
perturbations on its orbit from the other terrestrial planets. These large-amplitude oscillations drive large-scale variations in its
seasonal cycles of water, carbon dioxide and dust (e.g. Kieffer \& Zent, 1992). In addition, the large obliquity changes cause frequent
ice ages on Mars, during which there is a build-up of polar ice towards the equator (Head et al, 2003). The planet is currently in an
interglacial period having suffered an ice age approximately 1~Myr ago when the obliquity reached over 30$^\circ$ (Head et al, 2003).
Over the past 10~Myr the mean obliquity has decreased from approximately 35$^\circ$ to its current value of 25$^\circ$ (Touma \&
Wisdom, 1993; Laskar et al. 2002), indicating that past ice ages may have lasted much longer. \\

However, the conditions on Mars were very different shortly after its formation compared to what they are today. During the Noachian
period, which lasted from Mars' formation until 3.5~Gyr ago, there is evidence for a short existence of liquid water during its later
stages (e.g. Carr, 1996; Christensen et al., 2001), a temporary magnetic field (Lillis et al., 2008) and possibly a thick atmosphere
with the temperature at the surface exceeding 273~K. The existence of ubiquitous valley networks has led to the idea that fluvial
erosion was an important process on Noachian Mars (Carr, 1996). However, Christensen et al. (2001) argue that the duration of liquid
water flow must have been brief because of the existence of olivine and lack of carbonates at the surface.By the end of the Noachian,
the bulk of the valley networks had formed and the erosion rate was steeply declining (Carr, 1996). In addition, the magnetic field
might have protected the atmosphere from pick-up-ion sputtering and hydrodynamic collisions, and therefore knowledge of its evolution
is crucial for understanding the Noachian climate history. Isotopic evidence in Martian meteorite ALH84001 suggests that the atmosphere
was largely unfractionated near the end of the Noachian (Marti et al., 2001; Mathew \& Marti, 2001). The geologic and isotopic
information taken together suggests a relatively rapid loss of atmosphere as Mars entered the Hesperian era -- which lasted from
3.5~Gyr ago to 1.8~Gyr ago. Loss of a global magnetic field has been considered as an attractive mechanism, but this
probably happened much earlier (Lillis et al, 2008). \\

The end of the Noachian age corresponds approximately with the beginning of the Lunar Cataclysm or Late Heavy Bombardment (LHB). This
episode of intense cratering on the Moon (Tera et al., 1974), and the other terrestrial planets, occurring some 3.8~Gyr ago, formed
some of the basins on the Moon, notably Imbrium and Orientale. The cratering intensity was dramatically higher than what is predicted
from the remnants of planet formation (Bottke et al., 2007), suggesting a major re-ordering event in the solar system. Indeed, the LHB
is believed to be the last major event that sculpted the solar system (Strom et al., 2005). The LHB is also thought to be linked with
a dynamical instability in the outer solar system, which resulted in a reshuffling of giant planet orbits (Gomes et al., 2005). This
dynamical instability also affected the orbits of the terrestrial planets and probably created their current orbital structure (Brasser
et al., 2009). These models favour a fast migration of Jupiter and Saturn, resulting in a short-lived, high-eccentricity phase after a
long spell of circular orbits (Brasser et al., 2009). Recent results have uncovered that the orbits of the giant planets were likely
less dynamically excited (lower eccentricities and inclinations), and that they were much closer together (Morbidelli et al., 2007).
However, for this study we shall not limit ourselves to this one particular case, but explore a wider range of possible
configurations. Unfortunately, the orbits of the terrestrial planets before the LHB are unknown, but there is a range of possible
solutions. Therefore we create a set of different terrestrial planet systems with a wide range of initial conditions, and determine the
influence of their perturbations on the obliquity evolution of Mars using numerical simulations. The simulations are inspected for
cases where the obliquity could be stable on long time scales. \\

This paper proceeds as follows. Section~2 contains a short presentation of the equations of motion of the obliquity of Mars, and the
typical values of the quantities that enter these equations. Section~3 contains a description of our numerical methods. In Section~4
we present our results, and the last section contains our conclusions.

\section{Evolution of the obliquity with planetary perturbations}
\label{2}
The evolution of the obliquity of a planet subject to perturbations from other planets has been studied by various authors (e.g.
Colombo, 1966; Ward, 1974; Kinoshita, 1977; Laskar et al., 1993; Ward \& Hamilton, 2004).
Here we give a short summary based on the elegant method of N\'{e}ron de Surgy \& Laskar (1997). \\

Mars' figure exhibits precession of the equinoxes, similar to Earth. The precession is caused by Mars' figure being slightly
flattened, so that there is a normal component to the force exerted on Mars by the Sun. This component of the force induces a torque
on Mars that drives the precession and which acts in the direction opposite to Mars' orbital motion. The Hamiltonian describing the
torque on Mars is (N\'{e}ron de Surgy \& Laskar, 1997)

\begin{equation}
 \mathcal{H} = \frac{L^2}{2\mathcal{C}} - \frac{\alpha X^2}{2L},
\end{equation}
where $L=\mathcal{C}\omega_r$ is the rotational angular momentum of Mars, $\mathcal{C}$ is Mars' largest moment of inertia,
$\omega_r$ is its rotation rate, $X=L\cos \varepsilon$ with $\varepsilon$ being the obliquity. The precession of the equinox,
$\dot{\psi}$, is governed by the parameter $\alpha$ as $\dot{\psi}=\alpha X$ where

\begin{equation}
 \alpha = \frac{3n^2}{2\omega_r}\frac{J_2\mathcal{M}R_M^2}{\mathcal{C}}(1-e^2)^{-3/2}.
\end{equation}
Here $J_2$ is Mars' quadrupole moment, $n=\sqrt{GM_{\odot}/a^3}$ is Mars' orbital mean motion, $R_M$ is Mars' radius, $\mathcal{M}$ is
its mass, $a$ is the semi-major axis and $e$ is the eccentricity. For simplicity we will set $L=1$ and thus $X = \cos \varepsilon$. The
most recent data for the figure, obliquity, equinox and orbit of Mars are taken from Folkner et al. (1997), with $\varepsilon =
25.189517^\circ$, $\psi = 35.43777^\circ$, $\omega_r=350.89198521^\circ$/day, ${\mathcal{C}}/{\mathcal{M}}R_M^2=0.3662$, $J_2=1.96045
\times 10^{-3}$, $a=1.523698$~AU and $e=0.0933428$, from which we compute the current value $\alpha = 8.373$~$\arcsec$/yr and thus
$\alpha \cos \varepsilon =7.576$~$\arcsec$/yr. \\

However, the orbit of Mars is not static but undergoes periodic changes induced by perturbations from the other planets. When taking
these perturbations into account, the ecliptic is no longer an inertial plane and thus the kinetic energy of its forcing,
$\mathcal{E}$, has to be added to the Hamiltonian (Kinoshita, 1977). This is given by

\begin{equation}
\mathcal{E} = 2\Gamma(t)X-\sqrt{1-X^2}[A(t)\sin \psi + B(t) \cos \psi],
\end{equation}
where

\begin{equation}
\Gamma(t) = \dot{p}q-\dot{q}p; \quad A(t)= 2\frac{\dot{q}+p\Gamma(t)}{\sqrt{1-p^2-q^2}}; \quad
B(t)= 2\frac{\dot{p}-q\Gamma(t)}{\sqrt{1-p^2-q^2}}
\end{equation}
with $\zeta=q+{{\rm{i}}}p = \sin(i/2)\exp({{\rm{i}}}\Omega)$ and i$^2=-1$. Here $i$ is the inclination of the orbit of Mars with
respect to the invariable plane of the solar system and $\Omega$ is its longitude of the ascending node. The new Hamiltonian is then
$\mathcal{K}=\mathcal{H}+\mathcal{E}$, and Hamilton's equations are

\begin{equation}
\label{spineq}
\dot{X}= \frac{\partial \mathcal{K}}{\partial \psi} = -\sqrt{1-X^2}[A(t) \cos \psi-B(t)\sin \psi]; \quad
\dot{\psi}= -\frac{\partial \mathcal{K}}{\partial X} = \alpha X -
X(1-X^2)^{-1/2}[A(t)\sin \psi +B(t) \cos \psi]-2\Gamma(t).
\end{equation}
Here $A(t)$, $B(t)$ and $\Gamma(t)$ are the forcing terms from the planetary perturbations on the orbit of Mars. Equations
(\ref{spineq}) are singular when $X=1$ i.e. when $\varepsilon = 0$, so for numerical integrations we follow Laskar et al. (1993) and
use $\chi =\xi +{{\rm{i}}}\eta= \sin \varepsilon\,\exp({{\rm{i}}}\psi)$, which yields

\begin{equation}
\dot{\xi} = A(t)\sqrt{1-\xi^2-\eta^2} -\eta[\alpha \sqrt{1-\xi^2-\eta^2} -2\Gamma(t)]; \quad \dot{\eta} =
-B(t)\sqrt{1-\xi^2-\eta^2} +\xi[\alpha \sqrt{1-\xi^2-\eta^2} -2\Gamma(t)].
\label{spineqreg}
\end{equation}
Equations (\ref{spineqreg}) contain no singularities. Now that we have outlined the basic equations of motion of the spin axis, we
turn to our numerical methods.

\section{Numerical Methods}
\label{3}
Unlike Laskar et al. (2004), our aim is not to obtain an accurate prediction of the future evolution of the obliquity of Mars, but
rather to test its stability as a function of varying initial conditions: different orbits of the terrestrial planets and a range of
original obliquities. Thus we integrate the equations of motion of the obliquity of Mars (\ref{spineqreg}) from existing simulations of
the solar system. This is described below.

\begin{itemize}
 \item We perform simulations of the solar system with various initial conditions of the giant planets and the terrestrial planets
(see next section) for 265 Myr with a time step of 0.01 yr (3.6525 days). The data is output every 1000~yr. We use the 2nd order MVS
integrator of Laskar \& Robutel (2001), implemented (Bro\v{z} et al., 2005) in the SWIFT integration package (Levison \& Duncan, 1994).
The code includes digital filtering of the orbital elements using Kaiser windows (Kaiser, 1966), based on the method of Quinn et
al. (1991). Any signal with a period shorter than 667~yr is suppressed. To mimic the effect of general relativity we added a
correction term as described in Nobili \& Will (1986). 

\item From the simulation the quantities $\alpha$, $A(t)$, $B(t)$ and $\Gamma(t)$ are calculated using the filtered elements.
Numerical forward differentiation with four data points was used to compute $\dot{p}$ and $\dot{q}$.

\item The quantities $\alpha$, $A(t)$, $B(t)$ and $\Gamma(t)$ are used to integrate eqs.~(\ref{spineqreg}) with the Bulirsch-Stoer
method (Bulirsch \& Stoer, 1966) with a time step of 1000 years. For the intermediate steps that the integrator performs before
extrapolation to zero step size the corresponding intermediate values of $\alpha$, $A(t)$, $B(t)$ and $\Gamma(t)$ are computed using
linear interpolation. The input are the original values of $\varepsilon$ and $\psi$. The output is Fourier analysed according to the
method of Laskar (1988). The maximum and minimum values of $\varepsilon$ and the averaged proper rate of precession,
$\dot{\psi}$, are recorded.
\end{itemize}
The last two steps are repeated for different values of original $\varepsilon$. Generally $\varepsilon$ was varied between 0 and
90$^\circ$ with steps of $0.2^\circ$, while the original value of $\psi$ was kept at its current value. All steps were repeated for
the different configurations of the terrestrial planets.

\section{Results}
\label{4}
\subsection{Current solar system}
In order to be able to distinguish the obliquity evolution of Mars during the Noachian from the current one, we give a short summary of
the latter. For more details see Laskar \& Robutel (1993), Touma \& Wisdom (1993) and Laskar et al., (2004). \\

\begin{figure}
\resizebox{\hsize}{!}{\includegraphics[angle=-90]{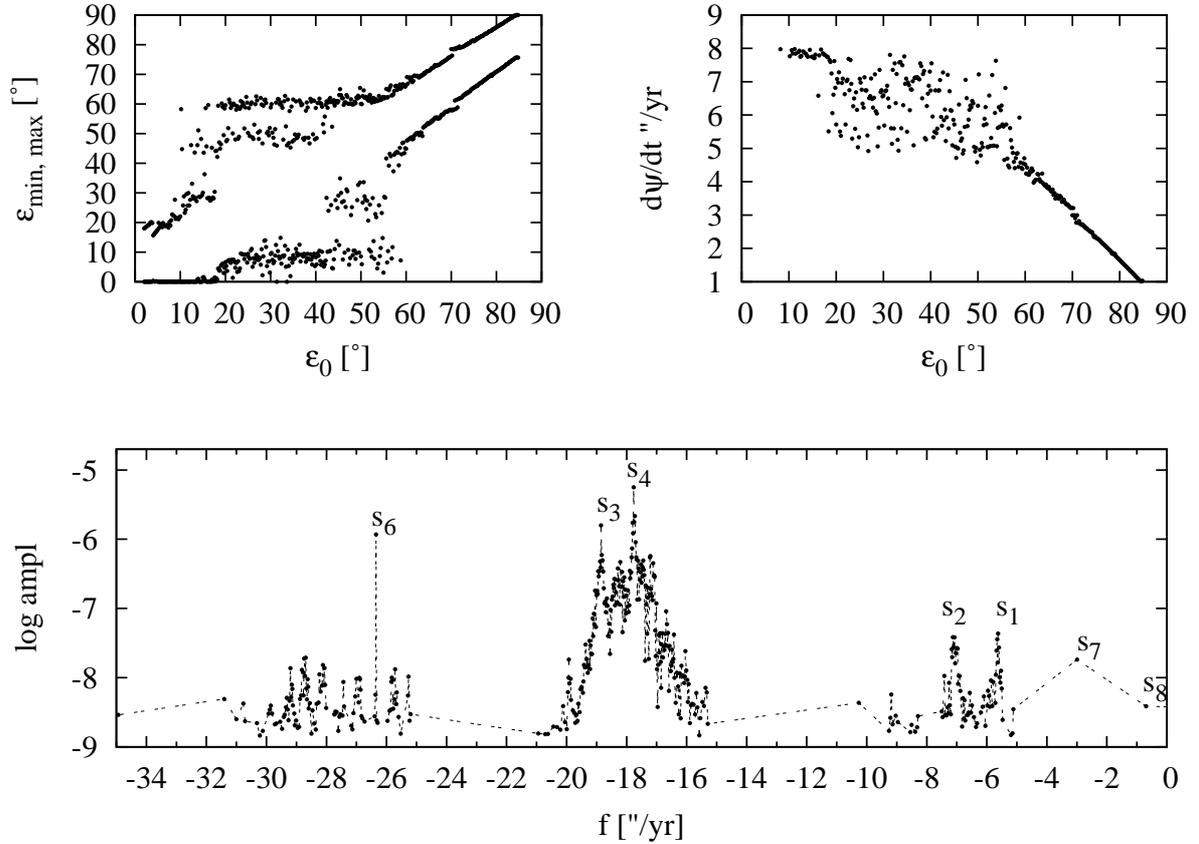}}
\caption{Top left panel shows maximum and minimum values of obliquity evolution (vertical axis) taken over 256~Myr as a function of the
original value (horizontal axis). Top right panel shows the averaged proper precession rate, $\dot{\psi}$ (vertical axis) vs original
obliquity. The bottom panel shows a spectrum of $A(t)+{\rm{i}}B(t)$, taken over 64~Myr. The vertical axis corresponds to the log of the
amplitude of each term while the horizontal axis corresponds to the frequency in arcsec per year. This figure is similar to those
presented in Laskar \& Robutel (1993).}
\label{oblcurr}
\end{figure}

Fig.~\ref{oblcurr} depicts current obliquity evolution, and should be compared directly with those presented in Laskar \& Robutel
(1993). The top left panel of Fig.~\ref{oblcurr} depicts the minimum and maximum values that the obliquity reaches during 256~Myr
(vertical axis) as a function of initial obliquity, $\varepsilon_0$ (horizontal axis). The top right panel depicts the average proper
precession rate of the equinoxes, $\dot{\psi}$, on the vertical axis as a function of $\varepsilon_0$. There are three regions of
motion. The first is for initial obliquities $<10^\circ$, where it oscillates between 0 and 20$^\circ$ and the precession decreases
with the obliquity as $\dot{\psi} = \alpha \cos \varepsilon$. The second region is between 10$^\circ$ and 60$^\circ$. Here the minimum
and maximum obliquities are 10$^\circ$ and 60$^\circ$ respectively and $\dot{\psi}$ ranges from 5~$\arcsec$/yr to 8~$\arcsec$/yr
without a trend with $\varepsilon$. This is the chaotic region identified by Laskar \& Robutel (1993) and Touma \& Wisdom (1993). On
Gyr time scales, the mean obliquity of Mars will vary between 0 and 60$^\circ$. Above 60$^\circ$ the motion becomes regular. The bottom
panel depicts the Fourier spectrum of $A(t)+{\rm{i}}B(t)$, similar to Laskar \& Robutel (1993), but taken over 64~Myr. The vertical
axis lists the log of the amplitude of each term while the horizontal axis corresponds to the frequency in arcsec per year. The forcing
peaks from the solar system's nodal eigenfrequencies $s_1 \ldots s_8$, corresponding to the proper regression frequencies of the
nodes of Mercury ($s_1$) to Neptune ($s_8$), (Brouwer \& van Woerkom, 1950), are indicated.\\ 

It is important to distinguish the two sources of chaos in the current evolution of Mars' obliquity. The first is caused by a
resonance overlap between the precession of Mars' figure and two eigenfrequencies of the solar system. The other is a secular
resonance between the orbits of Mercury, Venus and Jupiter, which indirectly affects the orbit of Mars. We shall examine each of these
below. \\

Usually the relation $\dot{\psi} = \alpha \cos \varepsilon$ is a good approximation for the precession rate with increasing obliquity.
For Mars, however, as the obliquity increases there are two values where $\dot{\psi}=|s_i|$ i.e. a secular spin-orbit resonance occurs.
The first resonance is with $s_2$ at $\varepsilon_2 =\arccos(s_2/\alpha)=32.3^\circ$. The second is with $s_1$ at $\varepsilon_1 =
\arccos(s_1/\alpha) = 48.0^\circ$. The width of these two resonances is directly proportional to the forcing on Mars' orbit from
Mercury and Venus. In the current solar system the forcing from both planets is so large that these resonances overlap, which occurs at
$\varepsilon \sim 40^\circ$. Resonance overlap is a source of chaos (Chirikov, 1979), which accounts for the chaotic region between
initial obliquities of 30$^\circ$ and 60$^\circ$. However, there is a second source of chaos that affects the obliquity, which arises
from chaos in the orbit of Mars. \\

Laskar (1990) identified two sources of chaos in the orbits of the terrestrial planets. The one that affects Mars' obliquity has
the argument $\sigma=(g_1-g_5)-(s_1-s_2)$, which is currently in libration with period 10~Myr (Laskar, 1990). Here $g_1 \ldots g_8$
are the eigenfrequencies of the longitudes of pericentre, corresponding to the proper precession frequencies of the perihelia of
Mercury ($g_1$) to Neptune ($g_8$) (Brouwer \& Van Woerkom, 1950). The argument $\sigma$ stays in libration for the next 200~Myr, but
its amplitude changes, implying the motion is near a separatrix. Laskar (1990) showed that the components associated with $s_1$
and $s_2$ in Mars' orbit are accompanied by a number of smaller components of similar frequency. A multiplet of components can also be
viewed as a single component with varying frequency and amplitude, with the frequency of the multiplet being comparable to the
frequency spread of the multiplet. These multiplets are caused by the presence of the secular resonance $(g_1-g_5)-(s_1-s_2)$. Thus in
the Fourier spectrum of Mars' orbit the lines associated with $s_1$ and $s_2$ are not clean delta functions but exhibit a near-Gaussian
profile i.e. they have significant sidebands. This is clearly visible in fig.~\ref{oblcurr} and the profiles overlap. These mutiplets
effect the motion of the mean obliquity because $\dot{\psi}$ can resonate with all these additional frequencies in the multiplets. The
mean obliquity pops in and out of resonance because the location of the separatrix varies with time. The crossing of the separatrix
occurs at essentially random phase which causes the obliquity to increase or decrease at random, and thus the mean obliquity exhibits
something akin to a random walk. The secular resonance acts when the mean obliquity $<30^\circ$. This implies that the mean obliquity
is chaotic from 0$^\circ$ to 60$^\circ$ on Gyr time scales (Laskar \& Robutel, 1993; Laskar et al., 2004). \\

From the above it appears there are two ways the obliquity of Mars can be stabilised. The first is to reduce the forcing from
Mercury and Venus on Mars so that the two resonances no longer overlap. This can be achieved by reducing the eccentricities and
inclinations of the terrestrial planets. There is indication that the terrestrial planets were situated on dynamically colder orbits
before the LHB (Brasser et al, 2009). The second method is to change the precession frequencies of either Mercury, Venus or
Jupiter and break the resonance $(g_1-g_5)-(s_1-s_2)$. There are no indications that the terrestrial planets had their orbits changed
significantly, but there is plenty of evidence that Jupiter had migrated (e.g. Fern\'{a}ndez \& Ip, 1984; Malhotra, 1993; Tsiganis et
al., 2005; Morbidelli et al., 2007, 2009, 2010; Brasser et al., 2009). Displacing Jupiter would change the value of $g_5$ and if
increased enough it will break the secular resonance. The effect of removing this resonance on the obliquity of Mars, while keeping
everything else the same, is discussed in the next subsection. 

\subsection{Compact configuration of the giant planets}
Here we investigate the effect of a different configuration of the giant planets on the obliquity evolution of Mars. It was almost
certain that the giant planets had different orbits before the LHB, so it is essential to study the stability of Mars' obliquity using
other configurations of the giant planets. We look for a system that increases $g_5$ so that the resonance $(g_1-g_5)-(s_1-s_2)$ is not
in libration. \\

The value of $g_5$ depends mostly on the separation between Jupiter and Saturn i.e. on their period ratio $P_S/P_J$.
Brasser et al. (2009) showed that when $P_S/P_J>2$, $g_5$ decreases sharply as $P_S/P_J$ increases: from approximately 20$\arcsec$/yr
when $P_S/P_J=2.03$ to its current value of 4.26$\arcsec$/yr when $P_S/P_J = 2.48$. As $P_S/P_J$ increases and $g_5$ decreases, it
crosses the values of $g_2=7.45\arcsec$/yr and $g_1=5.56\arcsec$/yr when $P_S/P_J \sim 2.1$ and $P_S/P_J \sim 2.3$ respectively. In
order not to destabilise the terrestrial system we need $g_5>g_2$ and thus $P_S/P_J<2.1$. There are several well-studied models in the
literature that prefer an initial period ratio between Jupiter and Saturn $P_S/P_J<2.1$. These are the model of Malhotra (1993), the
Nice model of Tsiganis et al. (2005) and the resonant model of Morbidelli et al. (2007). Here we use the configuration of Malhotra
(1993), since it has the most relaxed initial conditions for the giant planets: the initial period ratio is $P_S/P_J=2.05$ and the
giant planets have their current inclinations and eccentricities. The terrestrial planets remained on their current orbits. \\ 


\begin{figure}
\resizebox{\hsize}{!}{\includegraphics[angle=-90]{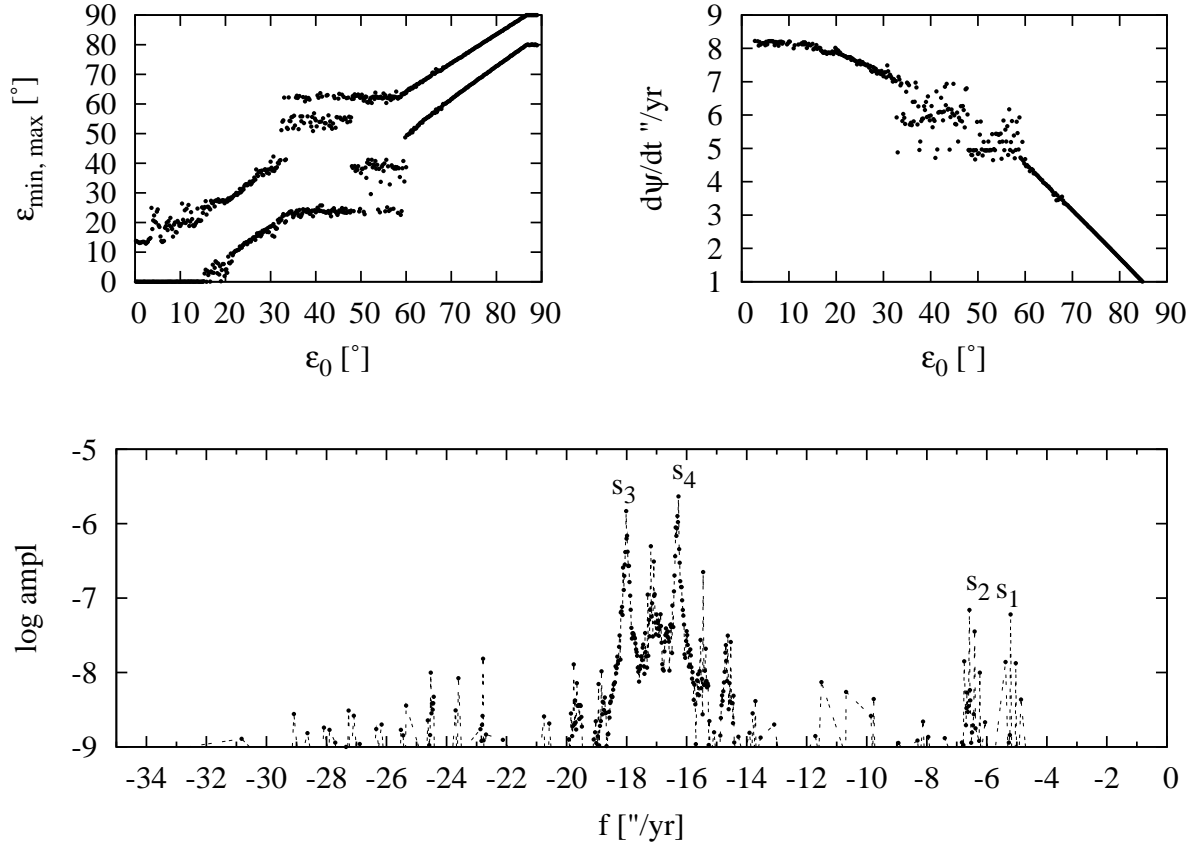}}
\caption{Similar to Fig.~\ref{oblcurr}, but now the giant planets are on the configuration of Malhotra (1993, 1995). The top-left panel
depicts the maximum and minimum obliquity as a function of original obliquity. The top-right panel shows the averaged proper precession
frequency. Both two panels are taken over 256~Myr. The bottom panel shows a spectrum of $A(t)+{\rm{i}}B(t)$, taken over 64~Myr. Notice
that the chaotic motion is confined to values of the initial obliquity between 35$^\circ$ and 60$^\circ$.}
\label{oblrenu}
\end{figure}

Figure~\ref{oblrenu} shows the result of numerical simulations using Malhotra's planets. The layout is exactly the same as
fig.~\ref{oblcurr}, and there are some significant differences. First the chaotic region is much narrower. In the current solar system
the chaotic region ranges from 10$^\circ$ to 60$^\circ$, though from 0 to 60$^\circ$ on longer time scale (Laskar \& Robutel, 1993;
Laskar et al., 2004). In fig.~\ref{oblrenu}, the chaotic region is confined to $\varepsilon_0$ from 35$^\circ$ to 60$^\circ$. The chaos
from the secular resonance $(g_1-g_5)-(s_1-s_2)$ has disappeared because $g_5 > g_2$. The only source of chaos is the overlap of the
two resonances $\dot{\psi}=|s_2|$ and $\dot{\psi}=|s_1|$. This can be inferred from the bottom panel of fig.~\ref{oblrenu}: the lines
around $s_1$ and $s_2$ are much narrower than in fig.~\ref{oblcurr}, implying a cleaner spectrum and thus a periodic signal instead of
a chaotic one. The obliquity can no longer resonate with all the intermediate frequencies and the chaos is confined to the main
resonances. By increasing $g_5$ and breaking the resonance $(g_1-g_5)-(s_1-s_2)$ the regions of initial obliquity with values below
35$^\circ$ and above 60$^\circ$ are stable, though the amplitude of oscillation is still large. This could be reduced if the
terrestrial planets were dynamically colder, which is discussed in the next subsection.

\subsection{Compact giant planets and colder terrestrial planets}
Here we examine the stability of the obliquity of Mars in a solar system where the orbits of the terrestrial planets are dynamically
colder.\\

The orbits of the terrestrial planets prior to the LHB cannot be obtained in similar fashion to those of the giant planets: the
former were formed by solid body accretion over 100~Myr time scales well after the gas disc dissipated (e.g. Raymond et al. 2009).
Their original orbital configuration was determined by the stochastic process of planetary embryo collision and accretion. 
Studies of the dynamical instability responsible for the LHB, likely caused by giant planet scattering which caused a large semi-major
axis jump for Jupiter (Brasser et al., 2009; Morbidelli et al., 2010), cannot determine the exact effect this would have on the orbits
of the terrestrial planets individually, so we study them as a system. One useful metric is the Angular Momentum Deficit (AMD; Laskar,
1997), which measures the deviation of the terrestrial system from circular and coplanar orbits, and is given by (Laskar, 1997)

\begin{equation}
 AMD = \frac{\sum_{n=1}^4 m_n\sqrt{a_n}(1-\sqrt{1-e_n^2}\cos i_n)}{\sum_{n=1}^4 m_n \sqrt{a_n}}
\end{equation}
where $m_n$ is the mass of each planet Mercury to Mars in solar masses, the semi-major axes $a_n$ are measured in AU and the
inclinations $i_n$ are with respect to the invariable plane. Its current value is AMD$_0=1.4043 \times 10^{-3}$ and the partitioning
among the planets is 35.8\%, 19.9\%, 20.1\% and 24.2\% for Mercury to Mars. This is close to the 25\% one would expect for each
planet. \\

The AMD is used as the independent variable to construct a set of terrestrial planet orbits with a range of dynamical excitation,
ranging from 10\% to 100\% of the current value (where the current AMD is equal to unity from now on) in ten equal steps. 
In order to test some low AMD values, we chose to place the giant planets on the resonant, circular configuration of Morbidelli
et al. (2007), to remove any forcing from the giants on the terrestrials. It is assumed that the dynamical instability that caused the
LHB would only excite the terrestrial planets, thus increasing their AMD to its current value. We have tested this assumption by
performing simulations in which we subjected the terrestrial planets on random orbits with pre-assigned AMD values to a
so-called 'jumping Jupiter' scenario (Brasser et al., 2009). We consistently find an AMD increase of $\sim$ 1 and that the increase is
systematic, suggesting that the primordial AMD was most likely lower than its current value. This can be argued as follows. During the
instability Jupiter suddenly appears on an orbit that is closer to the terrestrials and more eccentric, and remains there. This adds a
forcing to the eccentricities of the terrestrials, whose amplitude depends on the eccentricity of Jupiter and the semi-major axis ratio
between Jupiter and the terrestrials. Thus the eccentricities of the terrestrial planets, and thus the AMD, experience a systematic
increase. Therefore we take the upper limit of the primordial AMD equal to 1 and focus our discussion on values closer to 0. \\

For simplicity we kept the share of the AMD of each terrestrial planet the same as its current one. For the current orbits of the
terrestrial planets, with the exception of Mercury, $e \sim \sin i$. Therefore, the share of the AMD for each planet was partitioned
equally among its eccentricity and inclination. The other orbital parameters remained equal to their current values. \\

\begin{figure}
\resizebox{\hsize}{!}{\includegraphics[angle=-90]{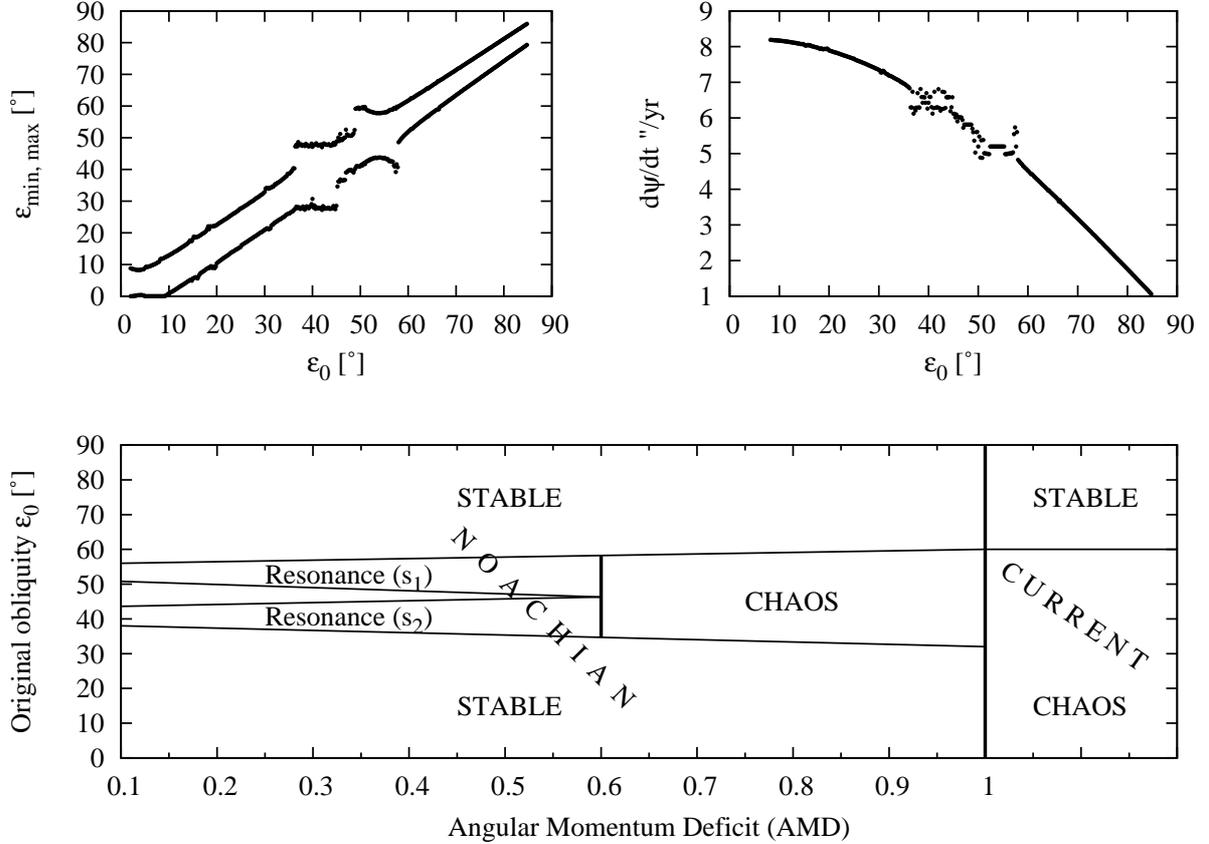}}
\caption{Similar to Fig.~\ref{oblcurr}, but now for a compact configuration of the giant planets and colder terrestrial planet orbits.
The AMD for the top two panels is equal to 40\% of the current value. The top-left panel depicts the maximum and minimum obliquity as a
function of original obliquity. The top-right panel shows the averaged proper precession frequency. The bottom panel is a summary that
shows the stable, chaotic and resonant regions as a function of initial AMD and initial obliquity for the systems studied here and the
current solar system.}
\label{oblsum}
\end{figure}

The summary of these numerical experiments is presented in Fig.~\ref{oblsum}, taken over 256~Myr. For the top two panels the AMD is
40\% of the current value. Note the small amplitude of oscillation of the obliquity, $\sim 12^\circ$, and strictly confined range of
obliquity motion for $\varepsilon_0 \lesssim 35^\circ$ in the top-right panel. This is an indication of stable motion. The first
resonant zone, where $\dot{\psi} = |s_2|$, occurs for initial obliquities 37$^\circ$ to 45$^\circ$. In resonance the obliquity
oscillates between 27$^\circ$ and 47$^\circ$ with a period of approximately 3~Myr. The second resonant zone, where
$\dot{\psi}=|s_1|$, is confined between initial obliquities of 48$^\circ$ to 56$^\circ$. Here the obliquity oscillates between
42$^\circ$ and 57$^\circ$ and with a period of 4~Myr. Above an initial obliquity of 57$^\circ$ the motion is stable yet again, and the
range of oscillation is merely $\sim 9^\circ$. The top-right panel depicts $\dot{\psi}$. There are a few very narrow chaotic zones
associated with the crossing of the separatrix between resonant and non-resonant motion. Inside the resonances $\dot{\psi}$ is a
constant and outside the resonates it varies as $\dot{\psi}=\alpha \cos \varepsilon$. \\

The bottom panel is a summary plot that shows the stable, chaotic and resonant regions as a function of initial AMD and initial
obliquity. The situation for the current solar system has been added to the right. During the Noachian there are two large stable
regions for mean obliquities below 35$^\circ$ or above 60$^\circ$. In between these regions the motion depends on the AMD: for values
less than 0.6 there are two stable resonant zones. These resonances start to overlap when the AMD is larger than 0.6 and the motion
becomes chaotic. The chaos is confined to the resonant zones because there is no other source of chaos. We expect the above results to
also apply when taking the giant planet configurations of Malhotra (1993) and Tsiganis et al. (2005). In all three models $g_5 > g_2$
so that the resonance $(g_1-g_5)-(s_1-s_2)$ has disappeared and thus the chaos is confined to the resonant region i.e. for
$\varepsilon_0$ between 30$^\circ$ and 60$^\circ$. For Malhotra's model, however, forcing from Jupiter limits the minimum AMD of the
terrestrial planets to approximately 30\% of the current one, while for the other two models the minimum AMD can reach all the way to
0.\\

Currently Mars' obliquity oscillates with an amplitude of approximately 20$^\circ$ and period 125~kyr (e.g. Ward, 1974). For the
systems studied in Fig.~\ref{oblsum} above, the obliquity oscillation amplitude in the lowermost stable region is also 20$^\circ$ when
the AMD is equal to the current value. The oscillation amplitude then decreases as AMD$^{1/2}$, so that it is just 6$^\circ$ when the
AMD is 0.1.

\section{Summary and conclusions}
\label{5}
We have analysed the evolution of the Martian obliquity for different initial conditions of both the terrestrial planets and the giant
planets, with a focus on the compact configuration of the giant planets presented in Morbidelli et al. (2007). The zones of stable
obliquity are summarised in the bottom panel of Fig.~\ref{oblsum}. \\

The obliquity could have been chaotic if the AMD of the terrestrial planets was more than 60\% of its current value, but only when
the mean value of the obliquity ranged from 30$^\circ$ to 60$^\circ$. For these values of the AMD, the secular spin-orbit resonances
$\dot{\psi}=|s_2|$ and $\dot{\psi}=|s_1|$ start to overlap. There, significant changes in the mean obliquity as well as large-range
obliquity oscillations become possible. Outside the resonant zones the obliquity oscillates with an amplitude set mostly by Mars' share
of and the total value of the AMD; the maximum value of the oscillation amplitude of the obliquity is approximately 20$^\circ$. Before
the LHB, the stable zone with obliquities lower than the resonant values is increased by approximately 25$^\circ$ when compared to the
current solar system. Therefore the obliquity was stable for any value of the AMD as long as its mean value was below $30^\circ$ or
above $60^\circ$. For values of the AMD less than 0.6, the obliquity is stable in the resonances with $s_2$ and $s_1$, but could suffer
large-amplitude, long-period oscillations.\\

{\bf{Acknowledgements}} \\
{\footnotesize RB and KW thank Germany's Helmholtz Alliance through their 'Planetary Evolution and Life' programme for financial
support. We also thank Alessandro Morbidelli and two anonymous reviewers for helpful suggestions.}

\section{References}
{\footnotesize
Bottke, W.~F., Levison, H.~F., Nesvorn{\'y}, D., Dones, L.\ 2007.\ Can planetesimals left over from 
terrestrial planet formation produce the lunar Late Heavy Bombardment?\ Icarus 190, 203-223. \\
Brasser, R., Morbidelli, A., Gomes, R., Tsiganis, K., Levison, H.~F.\ 2009.\ Constructing the secular architecture of the solar system
II: the terrestrial planets.\ Astronomy and Astrophysics 507, 1053-1065. \\
Brouwer, D., van Woerkom, A. J. J. 1950. The secular variations of the orbital elements of the principal planets. Astron. Papers
Amer. Ephem. 13, 81-107. \\
Bro{\v z}, M., Vokrouhlick{\'y}, D., Roig, F., Nesvorn{\'y}, D., Bottke, W.~F., Morbidelli, A.\ 2005.\ Yarkovsky origin of the unstable
asteroids in the 2/1 mean motion resonance with Jupiter.\ Monthly Notices of the Royal Astronomical Society 359, 1437-1455. \\
Bulirsch, R., Stoer, J. 1966. \ Numerical treatment of ordinary differential equations by extrapolation methods. \ Numerische
Mathematik 8, 1-13. \\
Carr, M.~H.\ 1996.\ Water on Mars.\ New York: Oxford University Press, 1996 . \\
Chirikov, B.~V. \ 1979. \ A universal instability of many-dimensional oscillator systems. Physics Reports 52, 263-379. \\
Christensen, P.~R., Morris, R.~V., Lane, M.~D., Bandfield, J.~L., Malin, M.~C.\ 2001.\ Global mapping of Martian hematite mineral
deposits: Remnants of water-driven processes on early Mars.\ Journal of Geophysical Research 106, 23873-23886. \\
Colombo, G.\ 1966.\ Cassini's second and third laws.\ The Astronomical Journal 71, 891. \\
Fernandez, J.~A., Ip, W.-H.\ 1984.\ Some dynamical aspects of the accretion of Uranus and Neptune - The exchange of orbital angular
momentum with planetesimals.\ Icarus 58, 109-120. \\
Folkner, W.~M., Yoder, C.~F., Yuan, D.~N., Standish, E.~M., Preston, R.~A.\ 1997.\ Interior Structure and Seasonal Mass Redistribution
of Mars from Radio Tracking of Mars Pathfinder.\ Science 278, 1749. \\
Gomes, R., Levison, H.~F., Tsiganis, K., Morbidelli, A.\ 2005.\ Origin of the cataclysmic Late 
Heavy Bombardment period of the terrestrial planets.\ Nature 435, 466-469. \\
Head, J.~W., Mustard, J.~F., Kreslavsky, M.~A., Milliken, R.~E., Marchant, D.~R.\ 2003.\ Recent 
ice ages on Mars.\ Nature 426, 797-802. \\
Jakosky, B.~M., Carr, M.~H.\ 1985.\ Possible precipitation of ice at low latitudes of Mars during periods of high obliquity.\ Nature 315,
559-561. \\
Jakosky, B.~M., Henderson, B.~G., Mellon, M.~T.\ 1995.\ Chaotic obliquity and the nature of the Martian climate.\ Journal of Geophysical
Research 100, 1579-1584. \\
Kaiser, J.~F.\ 1966. Digital Filters. In Kuo, F.~F. and Kaiser, J.~F. (Eds.), System Analysis by Digital Computer. New York,
Wiley. \\
Kieffer, H.~H., Zent, A.~P.\ 1992.\ Quasi-periodic climate change on Mars.\ Mars 1180-1218. \\
Kinoshita, H.\ 1977.\ Theory of the rotation of the rigid earth.\ Celestial Mechanics 15, 277-326. \\
Laskar, J.\ 1988.\ Secular evolution of the solar system over 10 million years.\ Astronomy and Astrophysics 198, 341-362. \\
Laskar, J.\ 1990.\ The chaotic motion of the solar system - A numerical estimate of the size of the chaotic zones.\ Icarus 88, 266-291.
\\
Laskar, J., Joutel, F., Boudin, F.\ 1993.\ Orbital, precessional, and insolation quantities for the Earth from -20 MYR to +10 MYR.\
Astronomy and Astrophysics 270, 522-533. \\
Laskar, J., Robutel, P.\ 1993.\ The chaotic obliquity of the planets.\ Nature 361, 608-612. \\
Laskar, J.\ 1997.\ Large scale chaos and the spacing of the inner planets..\ Astronomy and Astrophysics 317, L75-L78. \\
Laskar, J., Robutel, P.\ 2001.\ High order symplectic integrators for perturbed Hamiltonian systems.\ Celestial Mechanics and Dynamical
Astronomy 80, 39-62. \\
Laskar, J., Levrard, B., Mustard, J.~F.\ 2002.\ Orbital forcing of the martian polar layered 
deposits.\ Nature 419, 375-377. \\
Laskar, J., Correia, A.~C.~M., Gastineau, M., Joutel, F., Levrard, B., Robutel, P.\ 2004.\ Long 
term evolution and chaotic diffusion of the insolation quantities of Mars.\ Icarus 170, 343-364. \\
Levison, H.~F., Duncan, M.~J.\ 1994.\ The long-term dynamical behavior of short-period comets.\ Icarus 108, 18-36. \\
Lillis, R.~J., Frey, H.~V., Manga, M.\ 2008.\ Rapid decrease in Martian crustal magnetization in the Noachian era: Implications for the
dynamo and climate of early Mars.\ Geophysical Research Letters 35, 14203. \\
Malhotra, R.\ 1993.\ The origin of Pluto's peculiar orbit.\ Nature 365, 819-821. \\
Marti, K., Marty, B., Mathew, K.~J.\ 2001.\ Martian Interior Volatiles: Indigenous Signatures and Early Evolution.\ Meteoritics and
Planetary Science Supplement 36, 122. \\
Mathew, K.~J., Marti, K.\ 2001.\ Early evolution of Martian volatiles: Nitrogen and noble gas 
components in ALH84001 and Chassigny.\ Journal of Geophysical Research 106, 1401-1422. \\
Morbidelli, A., Tsiganis, K., Crida, A., Levison, H.~F., Gomes, R.\ 2007.\ Dynamics of the Giant Planets of the Solar System in the
Gaseous Protoplanetary Disk and Their Relationship to the Current Orbital Architecture.\ The Astronomical Journal 134, 1790-1798. \\
Morbidelli, A., Brasser, R., Gomes, R., Levison, H.~F., Tsiganis, K.\ 2010.\ Evidence from the Asteroid Belt for a Violent Past
Evolution of Jupiter's Orbit.\ The Astronomical Journal 140, 1391-1401. \\
N\'{e}ron de Surgy, O., Laskar, J.\ 1997.\ On the long term evolution of the spin of the Earth..\ Astronomy and Astrophysics 318,
975-989. \\
Nobili, A.~M., Will, C.~M.\ 1986.\ The real value of Mercury's perihelion advance.\ Nature 320, 39-41. \\
Quinn, T.~R., Tremaine, S., Duncan, M.\ 1991.\ A three million year integration of the earth's orbit.\ The Astronomical Journal 101,
2287-2305. \\
Raymond, S.~N., O'Brien, D.~P., Morbidelli, A., Kaib, N.~A.\ 2009.\ Building the terrestrial planets: Constrained accretion in the
inner Solar System.\ Icarus 203, 644-662. \\
Strom, R.~G., Malhotra, R., Ito, T., Yoshida, F., Kring, D.~A.\ 2005.\ The Origin of Planetary Impactors in the Inner Solar System.\
Science
309, 1847-1850. \\
Tera, F., Papanastassiou, D.~A., Wasserburg, G.~J.\ 1974.\ Isotopic evidence for a terminal lunar cataclysm.\ Earth and Planetary Science
Letters 22, 1. \\
Touma, J., Wisdom, J.\ 1993.\ The chaotic obliquity of Mars.\ Science 259, 1294-1297. \\
Tsiganis, K., Gomes, R., Morbidelli, A., Levison, H.~F.\ 2005.\ Origin of the orbital architecture of the giant planets of the Solar
System.\ Nature 435, 459-461. \\
Ward, W.~R.\ 1974.\ Climatic variations on Mars. I. Astronomical theory of insolation.\ Journal of 
Geophysical Research 79, 3375-3386. \\
Ward, W.~R., Hamilton, D.~P.\ 2004.\ Tilting Saturn. I. Analytic Model.\ The 
Astronomical Journal 128, 2501-2509.}
\end{document}